\documentclass[screen,sigconf]{acmart}
\usepackage{subcaption}
\usepackage{caption}
\usepackage{dirtytalk}
\usepackage{graphicx}
\usepackage{wasysym}
\usepackage{colortbl}
\usepackage{booktabs}
\usepackage{multirow}
\usepackage{balance}

\AtBeginDocument{%
  \providecommand\BibTeX{{%
    \normalfont B\kern-0.5em{\scshape i\kern-0.25em b}\kern-0.8em\TeX}}}

\copyrightyear{2024} 
\acmYear{2024} 
\setcopyright{acmlicensed}\acmConference[CHI '24]{Proceedings of the CHI
Conference on Human Factors in Computing Systems}{May 11--16,
2024}{Honolulu, HI, USA}
\acmBooktitle{Proceedings of the CHI Conference on Human Factors in
Computing Systems (CHI '24), May 11--16, 2024, Honolulu, HI, USA}
\acmDOI{10.1145/3613904.3642787}
\acmISBN{979-8-4007-0330-0/24/05}

\begin{document}

\title[Third-Party Developers and Tool Development]{Third-Party Developers and Tool Development For Community Management on Live Streaming Platform Twitch}

\author{Jie Cai}
\affiliation{%
  \institution{Pennsylvania State University}
  \city{University Park}
  \country{USA}}
\email{jie.cai@psu.edu}
\orcid{0000-0002-0582-555X}

\author{Ya-Fang Lin}
\affiliation{%
  \institution{Pennsylvania State University}
  \city{University Park}
  \country{USA}}
\email{yml5563@psu.edu}
\orcid{0000-0003-3689-5910}
\author{He Zhang}
\affiliation{%
  \institution{Pennsylvania State University}
  \city{University Park}
  \country{USA}}
\email{hpz5211@psu.edu}
\orcid{0000-0002-8169-1653}
\author{John M. Carroll}
\affiliation{%
  \institution{Pennsylvania State University}
  \city{University Park}
  \country{USA}}
\email{jmc56@psu.edu}
\orcid{0000-0001-5189-337X}

\renewcommand{\shortauthors}{Cai et al.}

\begin{abstract}
Community management is critical for stakeholders to collaboratively build and sustain communities with socio-technical support. However, most of the existing research has mainly focused on the community members and the platform, with little attention given to the developers who act as intermediaries between the platform and community members and develop tools to support community management.  This study focuses on third-party developers (TPDs) for the live streaming platform Twitch and explores their tool development practices. Using a mixed method with in-depth qualitative analysis, we found that TPDs maintain complex relationships with different stakeholders (streamers, viewers, platform, professional developers), and the multi-layered policy restricts their agency regarding idea innovation and tool development. We argue that HCI research should shift its focus from tool users to tool developers with regard to community management. We propose designs to support closer collaboration between TPDS and the platform and professional developers and streamline TPDs' development process with unified toolkits and policy documentation.
\end{abstract}



\maketitle
\section{Introduction}
The advanced technological infrastructure reduces the barrier to content creators and facilitates user-generated content (UGC) to reach users worldwide on many online platforms. The increasing number of content creators and the exponential increase of multimodal content inevitably cause user and content management on many platforms because not all users and UGC follow the terms of services \cite{uttarapong_harassment_2021,lyu_i_2024}. On the one hand, many platforms promote good and civil behaviors to flourish in diverse online communities; on the other hand, they curb destructive behaviors and punish violators breaking community rules.

Many platforms provide technical support to empower stakeholders to perform most of the management tasks, such as automated moderation bots to support volunteer moderators for Subreddit \cite{dosono_moderation_2019, chandrasekharan_crossmod_2019} and live streamers in Twitch channels \cite{seering_moderator_2019, wohn_volunteer_2019}. To meet the diverse needs of sub-communities, many platforms at the same time provide open-access application programming interfaces (APIs) for third-party developers (TPDs) to develop tools (e.g., bots, extensions, plugins) to help users manage their communities with specific needs and norms, creating a diverse and thriving market of third-party tools \cite{marketus_api_2023}. Additionally, existing platform-provided tools also receive critiques from users that they often lack accuracy, do not meet the needs of a specific group of users, and increase the workload of users to some extent (e.g., more time to set up, need to review cases to approve) \cite{uttarapong_harassment_2021, cai_hate_2023, jhaver_human-machine_2019}. Many users use third-party tools to supplement their community management \cite{cai_categorizing_2019}.

Community management and moderation have been broadly discussed in HCI and CSCW (e.g., \cite{wohn_how_2017, kiene_volunteer_2019, mandryk_combating_2023}). However, many HCI scholarships about community management tool design focus on tool users' perspectives and explore issues about how to moderate (automatically or manually) (e.g.,  \cite{hsieh_nip_2023, chandrasekharan_crossmod_2019, chandrasekharan_bag_2017}), when to intervene (proactive or reactive) (e.g., \cite{cai_moderation_2021,cai_after_2021, seering_shaping_2017}), and what actions to take (account suspension or content removal) (e.g., \cite{kou_punishment_2021,jhaver_did_2019}). Tool users' perspectives highlight the role and importance of tools in their community management; the complaints about the platform's tools also highlight the need to improve the existing tool design.  The tension between tool importance for end-users and tool complaints to platforms points to a research gap related to tool developers, especially third-party developers (TPDs), who work as mediators between end-users and platforms and develop diverse tools to support the platform's ecosystem.  These TPDs are usually active community members who engage with the communities, identify issues, and react faster to crisis management than the platform does \cite{cai_hate_2023,han_hate_2023}. TPDs' essential role in community management, in addition to the tension between users' needs and the drawbacks of existing tools, highlights the need to understand tool design from the TPD's perspective. In this study, we ask the following research question: 

\begin{itemize}
    \item What are the practices of TPDs regarding tool development for live-streaming community management? 
\end{itemize}

This study aims to understand their discussion about tool development practices and identify their challenges. We chose a semi-private TPD community on Discord and crawled their discussion with topic modeling and in-depth qualitative analysis.  We found that TPDs experience multi-layered policy regulation regarding tool development; they also work as mediators between the platform and the tool users to create a management tool ecosystem but face challenges from the platform regarding technical, social, and policy-related issues. In line with HCI's work to build safe video-sharing platforms \cite{niu_building_2023,bartolome_literature_2023}, this study contributes to existing research about online community management with a bottom-up governance structure from TPDs' perspectives. We highlight how multi-layered policy and stakeholder relationships affect TPDs' development process with implications to support their autonomy and agency.

\section{Related Work}
\subsection{Tool Development For Online Community Management }

Managing communities to attract and integrate new users is essential for community growth.  New users play a critical role in the development and sustainability of online communities because of their potential to introduce fresh ideas and perspectives and contribute to the diversity of content \cite{kraut_challenges_2012}.  These new users can be temporal or regular help seekers or experts who can support others \cite{singh_types_2011}. As existing members may leave a community for various reasons, attracting and integrating new members becomes essential for maintaining an active and vibrant community. Tools like bots and extensions play social and functional roles to augment community administrators' management.

Maintaining communities requires deliberate efforts from the community and its members. This thread focuses on, on the one hand, how to facilitate community building and engagement of community members, such as designing chatbot and CAPTCHAs to facilitate community discussion \cite{kim_moderator_2021,seering_designing_2019},  setting a good example for the rest of the community members to mimic normative behaviors \cite{seering_shaping_2017, cai_moderation_2021}, providing rewards for community members' contributions (e.g., Karma point on Reddit), and showing recognition of loyal community members (e.g., Twitch badge system) \cite{cai_understanding_2023}. In this thread, bots play important social roles, such as greeting newcomers \cite{seering_social_2018},  encouraging participation {\cite{kim_bot_2020}}, and personalizing interactions and support \cite{yadav_feedpal_2019, rahman_adolescentbot_2021}.

While many HCI scholars focus on the community norms and policy \cite{chancellor_norms_2018, chandrasekharan_internets_2018} and algorithm design \cite{kou_mediating_2020,stringhini_evilcohort_2015,nilizadeh_poised_2017,kumar_community_2018, mariconti_you_2019} to improve the governance structure and performance. A thread of research focuses on the tool users,  such as general online viewers \cite{wright_recast_2021}, content creators \cite{jhaver_designing_2022}, and community administrators \cite{cai_after_2021,jhaver_human-machine_2019, kuo_unsung_2023}. This thread mainly focuses on bots'  functional roles \cite{zheng_roles_2019, smith_impact_2022},  such as comment-filter tools \cite{jhaver_designing_2022}, toxicity prediction tools \cite{wright_recast_2021}, evidence-capture tools \cite{sultana_unmochon_2021}, and harmful content blurring tools \cite{das_fast_2020}.

Though scholars have investigated different mechanisms to support community management and moderation from both the platform and tool users' perspectives, we know little about how TPDs implement these mechanisms and develop efficient tools for community management.

\subsection{Third-Party Developers and Collaboration}

TPDs are actors acting on behalf of the platform owner developing applications, services, or systems to address the needs of the end-users \cite{ghazawneh_balancing_2013}. Third-party development tends to be governed by the contractual relationship with the platform \cite{boudreau_how_2009}. TPDs are rarely directly compensated for their development work; instead, they are offered a marketplace for their applications with a greater reach to end-users \cite{west_browsing_2010}. They are a necessary part of the platform's ecosystem, collaborating with the platform to exchange innovative ideas and co-create commercial values \cite{qi_understanding_2021}. Platforms can diversify their services through a voluminous development in a wide range of application areas \cite{boudreau_let_2012}. By doing so, platforms can remain lean in terms of resources and labor on the platform while incorporating the ideas of the developers' communities in cultivating a digital ecosystem \cite{ghazawneh_balancing_2013}.  

To successfully build platform ecosystems, many platforms shift the focus from developing applications on their own to offering resources to facilitate TPDs in their development work \cite{prugl_learning_2006}, such as providing an interface between the platform and TPDs, to give these developers access to the core platform functionality to build applications. For example, the live streaming platform - Twitch provides an API for third-party developers to use part of the platform's functions to connect to tools they developed; these tools are faced by end-users such as streamers, moderators, and viewers.  Facilitating third-party development is a way for the platform to transfer its design capability to end-users and, in turn, generate complementary assets in the form of applications with their potential values \cite{teece_profiting_1986}.   

The existing HCI literature on developers focuses mainly on gaming developers \cite{freeman_exploring_2019,freeman_mitigating_2020,freeman_understanding_2023, freeman_pro-amateur-driven_2020, li_exploring_2023} or open-source software development, such as GitHub \cite{huang_effectiveness_2016}, and hardware development \cite{mellis_collaboration_2012}, with topics specifically focused on collaboration with support \cite{guzzi_supporting_2015, freeman_making_2016}, teamwork \cite{freeman_exploring_2019,freeman_tale_2021} and conflict in design \cite{filippova_mudslinging_2015}. For example, interaction designers and software developers face challenges in communicating preexisting interactions. There is a need to design collaborative tools to facilitate the designer's ability to express and the developer's ability to implement complex interactive systems \cite{maudet_design_2017}. 

Third-party development in the community management and moderation context is unique in two aspects: First, third-party developers are not developers within the organization regarding work status and viewpoints.  For example, professional and voluntary developers in open-source communities face collaborative challenges due to the different viewpoints and backgrounds of projects and programming \cite{van_wendel_de_joode_managing_2004}. Second, the community management and moderation context makes open source less possible since the moderation tools often target and can be potentially utilized by violators. It is risky to share the development process details with the public. These TPDs often have a (semi-)private space to discuss and share their knowledge. Considering the professionalism and the sensitive context, it is necessary to understand their practices to support them. In line with recent work to understand the TPDs' development process \cite{qi_understanding_2021}, this study would focus on TPDs and understand their tool development practices regarding streaming management and moderation and their roles in the platform ecosystem. 

\begin{figure*}
  \centering
  \includegraphics[width=.8\linewidth]{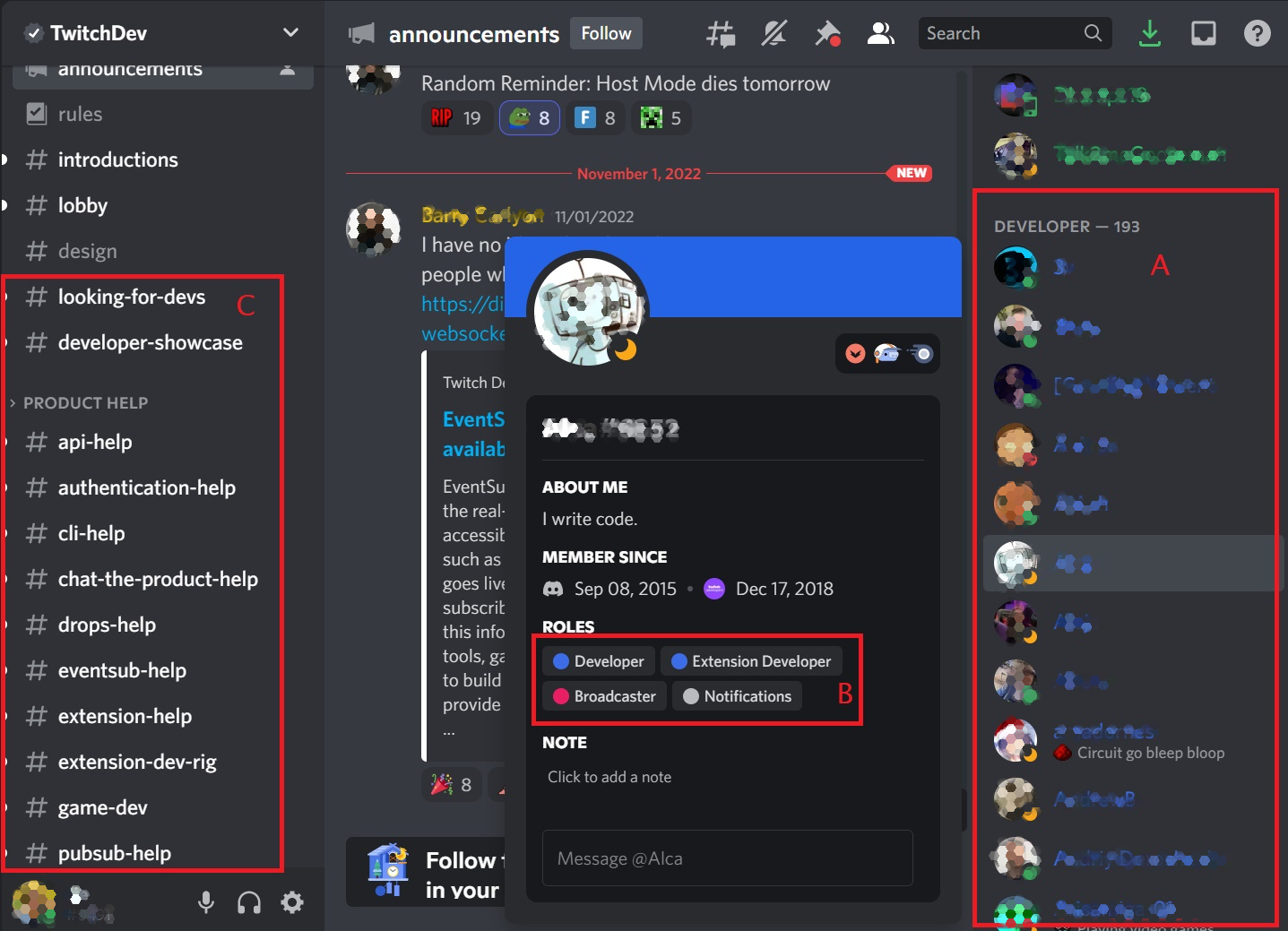}
  \caption{A screenshot of the \textit{TwitchDev} server on Discord. (A) the list of 193 active developers online; (B) a user with many different tagged roles verified by the server administrators such as \textbf{developer} and \textbf{extension developer}; (C) various channels for developers to discuss and collaborate on technical issues related to bot development.}
  \Description{A screenshot of the Twitch developer server on Discord. (A) the list of 193 active developers online; (B) a user with many different tagged roles verified by the server administrators such as \textbf{developer} and \textbf{extension developer}; (C) various channels for developers to discuss and collaborate on technical issues related to bot development.}
  \label{discord}
\end{figure*}

\subsection{Third-Party Developers on Discord Working for Twitch As the Research Context}

Live streaming as a new medium provides mass communication in the chatroom \cite{hamilton_streaming_2014}. While streamers are broadcasting with various low- and high-fidelity equipment \cite{drosos_design_2022}, viewers can register a pseudonymous account and send messages in the chatroom. The streamer can read and respond orally to these messages. Often, the streamer would like to interact with viewers to build communities \cite{sheng_virtual_2020} and promote prosocial behaviors in the chat \cite{seering_designing_2019}. As communities grow, streamers appoint some volunteer moderators and deploy bots to effectively manage the channels and viewership.

We choose Twitch as the research context because 1) live streaming is a promising industry with interactive and multimodal design and has thrived in decades and evolved with novel technology, 2) it is a leading live streaming platform in the industry, its other stakeholders (e.g., streamers, viewers, volunteer moderators) has caught much attention in HCI (e.g., {\cite{cai_coordination_2022,wohn_audience_2020, seering_shaping_2017, yen_storychat_2023}}). Who developed tools for these stakeholders receives little attention. Twitch has an extension and streaming tool market for third-party tools with different categories, such as an extension on the go, viewer engagement, stats and trackers, loyalty and recognition, etc. It also allows other third-party tools to connect to the platform, such as BetterTTV \footnote{\url{https://betterttv.com/}} and Frankerfacez \footnote{\url{https://www.frankerfacez.com/}}; these are all commonly used extensions on Twitch \cite{cai_categorizing_2019}. There are also other personally developed bots, such as Sery\_Bot \footnote{\url{https://serycodes.carrd.co/}}, which is a timely and valuable bot for streamers to combat human-bot coordinated attacks in live streaming communities \cite{han_hate_2023,cai_hate_2023}. All these tools are either partnered with Twitch or independently run by TPDs. 

Discord can be considered part of Twitch's ecology. Many streamers use Discord as a semi-private space for their community management \cite{cai_coordination_2022,cai_moderation_2021}. Unlike other public spaces allowing users to view content without registering or joining the community, such as Reddit, GitHub, and Stack Overflow, Discord only allows users who have joined the community, namely the server, to view its content and engage. All users can easily access and join the server with a link invitation. In the server, some public channels are visible to all users; some private channels are only available to a specific group of users. Discord allows users to communicate with text-based and voice-based chats. Initially, streamers usually use Discord to have group voice chat with teammates when playing and streaming video games. Off the stream, Discord also works as a space for streamers and their followers/fans to socialize. Nowadays, it works as a third space for people with similar interests. 

Since Twitch and Discord are closely connected, many TPDs for these two platforms have formed communities on Discord. The leading author joined the informal online communities called the \textit{TwitchDev} server. This server is run by non-staff volunteers such as moderators and administrators. TPDs will get a developer role in this Discord by the administrators of this Discord server if they prove that their program developments are building for Twitch users or using the Twitch-provided tools. The community has thousands of TPDs to share their experiences. It has more than 2000 active members daily, including Twitch official staff, TPDs, broadcasters, and viewers (as shown in \autoref{discord}). It includes many categories and channels, such as \textit{developer-showcase}, \textit{looking-for-devs}, and \textit{extension-help}. Usually, \textit{lobby} is the primary channel where everyone sends their first messages, asks questions, and socializes.

\section{Methods}
We focused on TPDs' tool development practices. We specifically refer to ``tool'' as ``bot and extension'' for community management, not systems or tools TPDs used to do development or programs they develop for Twitch or other platforms. We first crawled all the chat history of the Discord server.  After we went through all the channels, we realized that the ``lobby'' contains the most extensive dataset, and all other channels are comparatively small.  Then, we decided to focus on the ``lobby'' as the primary data source to run topic modeling to identify the bot-related topics. We use bot and extension development interchangeably based on TPDs' description. Next, we qualitatively analyze these topics to explore their practices and challenges.   

\subsection{Data Exploration and Pre-processing}\label{sec:data-exploration}
We collected 45,376 data entries from the largest channel, ``lobby.'' This data spans from October 25, 2018, to April 25, 2023. Each entry represents a message sent by a user in the channel and includes the content and timestamp of the message. Among these, we processed each message to remove English stopwords using the NLTK \footnote{\url{https://www.nltk.org/}} library. Combining our reading and reference to prior work with similar data structure \cite{ma_how_2021}, we excluded messages with fewer than ten words post-stopword removal to ensure a more meaningful and contextual analysis. This preprocessing step aligns with standard practices in text analysis to filter out potentially non-informative short texts ~\cite{tseng2007text}. We obtained 8,219 data entries, and this dataset comprised 279,581 words (1,567,282 characters). Based on our research questions and feasibility, the data collection from the channel was a one-time and fixed process. We did not consider any changes in the channel before or after the data collection.

\subsubsection{Topic Modeling}\label{subsec:topicmodeling}
\begin{figure*}[ht]
\centering
\begin{subfigure}{0.44\textwidth}
  \includegraphics[width=\linewidth]{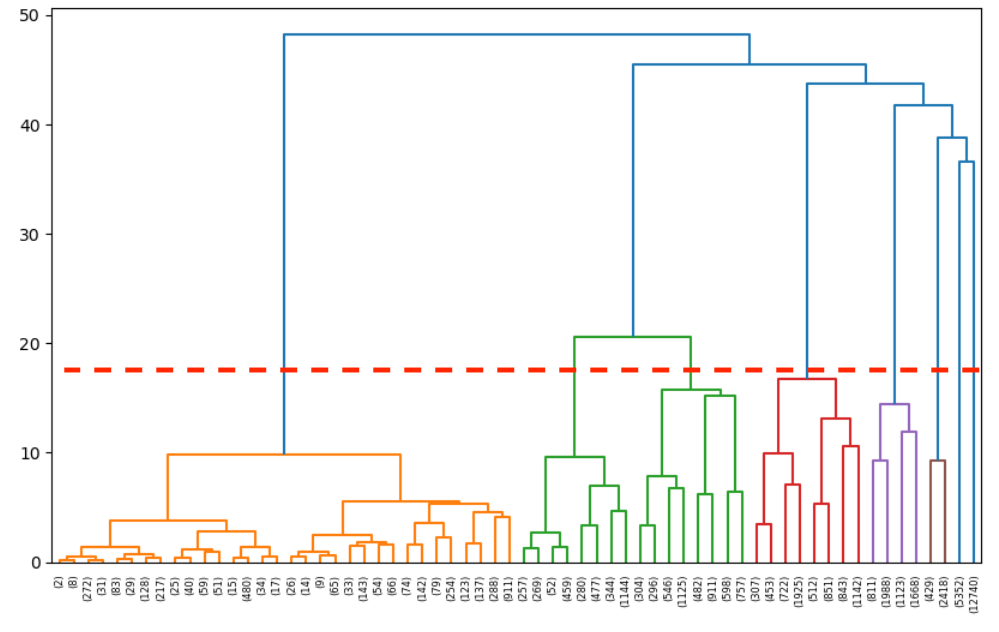}
  \caption{}
  \label{fig.Hierarchical-Clustering}
  \Description{Topic Modeling Results: (a) Result of Agglomerative Hierarchical Clustering Dendrogram. It displays the similarity between each cluster, where the red line effectively separates large and small clusters, making the size of each cluster after the division more reasonable. The clusters below the red line are more similar and more coherent than those above it. }
\end{subfigure}
\begin{subfigure}{0.55\textwidth}
  \includegraphics[width=\linewidth]{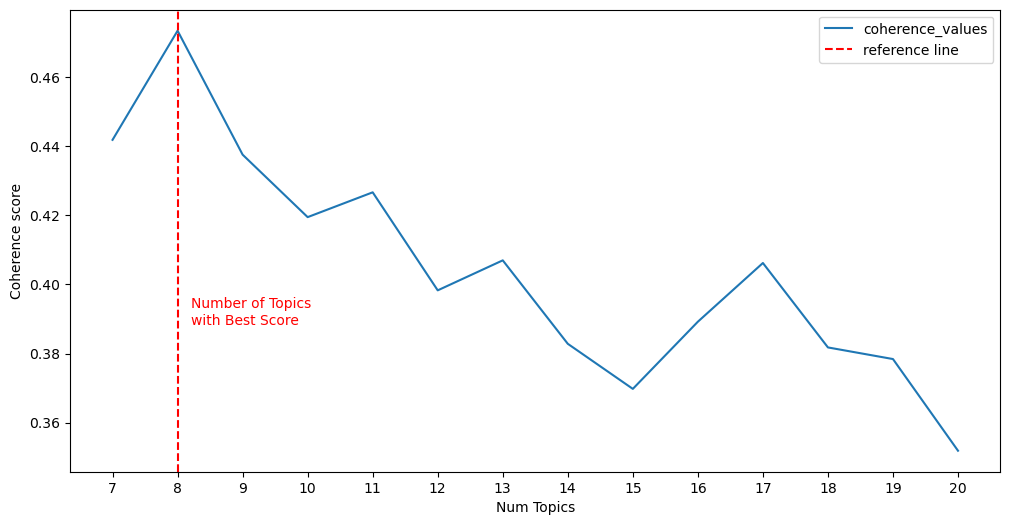}
  \caption{}
  \label{fig.coherencemodel}
    \Description{Topic Modeling Results:(b) Result of Coherence Model. It shows eight topics have the highest scores.}
\end{subfigure}
\caption{Topic Modeling Results: (a) Result of Agglomerative Hierarchical Clustering Dendrogram. The x-axis represents the clustering results of sample data, with numbers indicating the quantity of samples contained within each cluster. The y-axis represents the similarity during the merging of different clusters, where greater distance signifies lower similarity. (b) Result of Coherence Model. The x-axis represents the number of topics, and the y-axis represents the coherence score.}
    \label{}
\end{figure*}

Topic modeling facilitates the discovery of hidden thematic structures instead of relying solely on keyword retrieval {\cite{10.1145/2133806.2133826}}. Discord comments are short texts presenting sparse, noisy, and non-informative words, posing challenges in generating accurate topics. Prior research has highlighted the effectiveness of topic modeling in extracting thematic structures from such short-text datasets, making it a valuable method for discerning underlying themes and patterns \cite{5416713,likhitha2019detailed}.
We employ Latent Dirichlet Allocation (LDA)~\cite{blei2003latent} for topic modeling. The LDA, alpha, and beta parameters are set to the default value of 1.0 divided by the number of topics. We performed a cluster analysis on the pre-processed data and selected one potentially more dominant cluster quantity in the results (Figure~\ref{fig.Hierarchical-Clustering}), which inspired us in the number of thematic trends (8, red line). Coherence score is considered one of the most effective metrics for topic modeling~\cite{10.5555/2390948.2391052}. Hence, the number of topics is determined according to coherence score~\cite{stevens2012exploring} of the extracted topics given various possible topic numbers ranging from 7 to 20. Figure~\ref{fig.coherencemodel} shows that the consistency model suggests using eight topics as the optimal number. Table~\ref{tab:LDAcorpus} illustrates the computation extraction of topic distribution for generated results from using LDAvis~\cite{sievert2014ldavis}, where each result may be generated by multiple topics, and each topic contains several keywords.  

Because we aim to focus specifically on the subjects of ``bot'' and ``extension,'' we have chosen Topics 4 and 6 as the main candidates for further analysis. In the results of our topic modeling, Topic 4 prominently highlights ``extension'' as one of its pivotal terms, with a notable weight of 0.025, positioning it as the top term. This implies that discussions or narratives related to extensions are predominantly captured within this topic. Similarly, Topic 6 incorporates the term ``bots'' with a weight of 0.010, ranking it sixth, suggesting that it encapsulates discussions revolving around bots. Further scrutiny of the LDA topic modeling outcomes on the corpus reveals that besides Topics 4 and 6, other topics don't extensively cover contents related to ``bots'' and ``extension.'' While this doesn't rule out the presence of related content in other topics, focusing on Topics 4 and 6 can more efficiently and comprehensively delve into discussions regarding ``bots'' and ``extension'' within this channel. We organized the data labeled as Topics 4 and 6 in the topic modeling, and ultimately, we obtained 2,179 comments, constituting the ``corpus'' used in this study. 

We clarify that we focus on bots and extensions because they are closely related to the development we aim to discuss in this paper. We also quickly reviewed other topics and found they were about general information inquiries, specific tech support, and socialization among TPDs. These are related to the Twitch platform or their programming experience but not specifically related to third-party tool development.  For example, topics with ``UserVoice, Twitch, and API '' include limited comments on tool development. Still, these are eventually captured by the bot and extension topics, as we mentioned in the relevant findings sections.

\begin{table*}[t]
\resizebox{\textwidth}{!}{%
\begin{tabular}{l|rrrrrrrrrr}
\hline
 & Term 1 & Term 2 & Term 3 & Term 4 & Term 5 & Term 6 & Term 7 & Term 8 & Term 9 & Term 10 \\ \hline
Topic 0: & 0.043*"twitch" & 0.025*"help" & 0.020*"discord" & 0.019*"party" & 0.014*"com" & 0.010*"support" & 0.010*"account" & 0.010*"new" & 0.010*"third" & 0.010*"report" \\
Topic 1: & 0.022*"channel" & 0.017*"still" & 0.013*"would" & 0.013*"want" & 0.011*"people" & 0.010*"well" & 0.010*"name" & 0.009*"stream" & 0.009*"since" & 0.009*"channels" \\
Topic 2: & 0.022*"game" & 0.017*"twitch" & 0.012*"time" & 0.012*"see" & 0.012*"use" & 0.011*"example" & 0.010*"work" & 0.009*"thing" & 0.009*"uservoice" & 0.009*"guess" \\
Topic 3: & 0.039*"thanks" & 0.030*"answer" & 0.023*"http" & 0.020*"quest" & 0.015*"got" & 0.014*"answers" & 0.014*"join" & 0.012*"barry" & 0.010*"help" & 0.009*"language" \\
\rowcolor[HTML]{EFEFEF} 
Topic 4: & {\color[HTML]{CB0000} 0.025*"extension"} & 0.019*"chat" & 0.015*"twitch" & 0.015*"good" & 0.012*"message" & 0.011*"use" & 0.009*"code" & 0.009*"token" & 0.009*"need" & 0.009*"bits" \\
Topic 5: & 0.072*"twitch" & 0.030*"tv" & 0.019*"api" & 0.016*"contact" & 0.015*"ask" & 0.015*"dev" & 0.015*"user" & 0.012*"send" & 0.012*"luck" & 0.012*"know" \\
\rowcolor[HTML]{EFEFEF} 
Topic 6: & 0.018*"one" & 0.017*"use" & 0.016*"something" & 0.015*"thank" & 0.011*"already" & {\color[HTML]{CB0000} 0.010*"bots"} & 0.010*"wait" & 0.010*"go" & 0.009*"looking" & 0.009*"think" \\
Topic 7 & 0.033*"twitch" & 0.031*"support" & 0.030*"link" & 0.026*"us" & 0.022*"found" & 0.021*"looks" & 0.021*"help" & 0.019*"request" & 0.015*"please" & 0.015*"friends" \\ \hline
\end{tabular}%
}
\caption{The results of the LDA topic modeling analysis on the corpus. Each topic is represented by weighted terms (top 10 highest). The subtopics we focus on (``extension'' and ``bots'') are highlighted in red, and the topics we choose for further in-depth analysis are highlighted in gray.}
\label{tab:LDAcorpus}
\end{table*}

\subsection{Thematic Analysis} 

We undertook an exhaustive analysis of 2,179 comments (hereinafter referred to as the ``corpus'') pertinent to the topics of ``extension'' and ``bot.'' After completing the corpus compilation, it was disseminated among all authors for the coding process. We adopted a hybrid technique that combined inductive and deductive coding for theme extraction~\cite{fereday_demonstrating_2006}. Our coding process is based on the definitions by Corbin \& Strauss~\cite{service2009book} and is divided into three critical phases: Independent Coding (Open Coding), Codebook Development (Axial Coding), and Comprehensive Coding (Selective Coding).

During the Independent Coding phase, each author autonomously embarked on an initial, open-ended coding of a segment of the corpus. This approach ensured that every entry was evaluated without the predisposition of other viewpoints. We treated each entry as an independent observation. Each of the three primary coders was randomly assigned 300 comments from the corpus for individual assessment.
The Codebook Development phase followed, wherein all authors congregated to compare and consolidate their independent coding results. This step involved a meticulous comparison of individual codes, striving to establish a unified label for each comment based on collective consensus. For clarity, comments were sequentially labeled: for instance, the first comment was encoded as `1. Extension workshop', the second as `2. refresh before expiration of subbed topic', and so on. The objective was then to ascertain the applicability of these codes to subsequent comments. At this stage, coders try to find the relationships between different concepts and connect and organize the categories and concepts, and an embryonic codebook is formulated. The authors revisited the corpus with this preliminary framework to ensure coding uniformity. Discrepancies or ambiguities detected during this process were addressed and refined through iterative discussions. Multiple sessions culminated in finalizing the codebook, comprising 59 primary codes. For streamlined subsequent coding, two auxiliary codes, '0. irrelevant' and '60. not listed' were also integrated.

In the final Comprehensive Coding phase stage, three coders undertook independent coding of the residual data. Since the reliability will decrease as the number of coders and codes increase \cite{krippendorff_content_2004}, we acknowledge that we had a low coefficient (Fleiss’s kappa is .30) considering that we have three coders with a large codebook.  Instead, we changed our strategies to have hours-long meetings to review and discuss each other's coding with notes and revisions \cite{decuir-gunby_developing_2011}. Regular weekly meetings facilitated discussions on coding disparities, fostering alignment and consensus. During our meeting, we start by sharing our overall thoughts about the weekly coding updates. Then, we follow a protocol where a coder starts reading the comment and interpreting their code, followed by other coders interpreting it. Finally, we vote for one of the three codes or come up with an emerging code to better cover the core meaning of the comment. At the same time, coders exchange notes, which helps in generating topics and prospective categories. The coding process lasts for about six weeks, and each meeting lasts for about two hours. This iterative process, intertwined with functional codes, led to the categorization of codes and the formation of overarching themes. The trio of coders subsequently regrouped to critically assess and compare their coding interpretations, aiming for optimal consistency and precision. In instances of disagreements or ambiguities, collective discussions paved the way for resolutions. Conclusively, we pinpointed high-caliber comments for each category, fine-tuning their alignment to distill the final themes. Themes and sub-themes with codes are summarized in \autoref{apx:codebook}.

\subsection{Positionality Statement and Ethical Consideration}

Though we're not developers, our expertise and research interests align closely with this topic. All authors are online community researchers.  The lead author is a live streaming user and researcher and has observed this Discord community for two years. The other team members, also active Discord users, joined this particular Discord community months before data collection to understand TPDs'  discussion about their practices. This study analyzes publicly available comments without direct interaction with the content creators. This approach aligns with the guidelines set by the University Institutional Review Board (IRB), which does not classify it as ``human subjects research.'' In our commitment to upholding user privacy, we followed Internet research guidelines and have applied stringent anonymization measures to the research data \cite{franzke_internet_2020}. This involves the meticulous removal of any identifiable information, such as user IDs, ensuring the anonymity of the data sources and protecting the privacy of individuals involved. The data reporting includes only textual comments.

\section{Findings}
Discord provides a space for stakeholders to socialize, share information and resources, and have Q\&A conversations in general. Regarding their tool development practices, there are mainly four high-level themes: (1) TPDs shared positive experiences interacting with streamers and other TPDs to exchange and innovate design ideas; (2) the platform's technical updates caused uncertainty in their working schedule and extra labor for tool compatibility; (3) they also complained about the policy at the different levels restricting their developing agency with the intent to initialize changes; and (4) they appreciated the TPD communities providing social support and opportunities to learn from each other. Findings are summarized in \autoref{apx:findings}.

\subsection{Design Ideas Exchange and Innovation}

\subsubsection{Streamer's Need and Design Ideas}
TPDs talk with streamers in this discord server, and some of the TPDs are also streamers. Streamers mainly ask for advice or suggestions about extension developments that could fulfill the needs of their streaming, such as managing their channels, avoiding spammers and bots, and creating alternative tools that can edit their streaming interfaces easily for better customization or accessibility purposes. For example, a streamer wanted to create their stream editing tools to replace Streamlabs (a commonly used stream editing software).
\begin{quote}
    \say{Hi guys, I'm curious if it's possible to self-host a streamlabs lookalike. Basically my problem is streamlabs' html/css editor is pretty wonky, and their documentation isn't that great, I'd like to make my own one for my channel instead. Is it possible?}
\end{quote}

Many streamers want extensions that can provide rich interactivity to their audience. For example, some streamers want to be able to respond to their viewers promptly or interact with specific kinds of audiences, such as new followers; some want to enrich viewers' presentation ability; some seek flexible approaches of bit use; and some want to coordinate with bots more smoothly that viewers would not confuse broadcasters' and their bots' messages.
Though streamers already have plenty of extensions to use, they still reach out to TPDs for help because 1) they want to create customized features that are too personalized to find existing extensions; even if they try to develop extensions by themselves, they need technical advice to implement their ideas; 2) some streamers work with TPDs to develop specialized extensions, yet they need help when their collaborators are unavailable.

\subsubsection{Exchange Design Ideas Among TPDs}
TPDs shared various kinds of extension design ideas on the Discord server. One of the most shared extension categories is supporting streamers to entertain, manage, and interact with their viewers, such as “finding the most popular follower of a given channel,” “altering stream title,” “making a link to a specific channel on Twitch using their streamer id instead of username,” and ”extended the chat functionality.” 
Another commonly shared category is crisis management. TPDs shared bot designs that can prevent streamers from being attacked by malicious bots.
\begin{quote}
    \say{gotcha! ya, I think a bot that could identify with some reasonable certainty, users that are bots on the fly, would be *super* useful. it would also make the tool distinguishable from all of the other cross-channel shared list ban bots. just an idea}
\end{quote}

Some TPDs shared their sandbox version of their bots before publishing and collected feedback from their peers to advance the features and functions. For example, they would like to showcase their bots.
\begin{quote}
   \say{Made this extension that will save you a lot of time! Let me know what you think. It isn't live yet. It costs like 5 dollars to publish an extension, so I just wanted to see if people actually want something like this. [link to the showcase video of the extension]}
\end{quote}

Some TPDs might share their successful or failed experience, provide explanations, and exchange design ideas. After reading other people's design ideas, some TPDs provided their ideas as feedback. These ideas could be very different from their peers' ideas or an improvement built on the top of the proposed ideas. For example, when a TPD talked about extensions interacting with viewers, another TPD shared their idea that allows viewers to \textit{\say{create an incentive for the streamer to do a certain thing in a game}} by just-in-time crowdfunding with bits.
Furthermore, TPDs resonate with each other's designs and adapt them into their designs accordingly. As a TPD shared, \textit{\say{Sounds fun, i have something similar and started as a stream avatar clone I saw other streamers use and thought that was pretty cool to have so I wrote my own.}} They even enhanced the original design when making their own ones.  The review process may deny ideas related to legal and ethical issues (e.g., a crypto tipping extension). 

\subsection{Impact of Twitch Technical update}

\subsubsection{Schedule Uncertainty}
Some TPDs unfamiliar with the review process asked many questions about the update uncertainty. For example, a TPD team asked questions about the impact of a UI/UX update on the review process. 
\begin{quote}
    \say{So we've been working on a component extension for a bit, and have been building it for before the UI/UX update that takes effect on July 31st. We haven't submitted it yet, but are wondering how this deadline affects us because we'd like to move into the next step for testing by the 31st (white listing a streamer or two and having their entire audience be able to interact with it). What should we know if we don't submit by that deadline, or if we do and it's sent back for updating and we don't resubmit by the 31st. Just trying to cover all our bases.}
\end{quote}
In this case, the TPD team was wondering how a new UI/UX update would affect the evaluation of their extension and if their extension testing would be valid if they tested it before the UI/UX update. They might need to adjust their development schedule accordingly.

Sometimes, the review time was so long with no feedback, especially before a long break, as a TPD expressed, \textit{\say{i just hope they get to it before closing down the office till next year.}} 
They hoped the Discord community managers could personally contact the Twitch team to check what was going on since they had no feedback for months. 
\begin{quote}
    \say{It really is awesome that you guys are streaming etc... but we ([link to a third-party developed site]) are waiting already for 5 months (!!!!!!!!!) to get our extension finally approved. Can you pelase finally take care of it???? PLEASE!!!!!!!!}
\end{quote}

\subsubsection{Extra Labor Caused by Technical Updates}

TPDs think any technical update needs coordination with TPDs and considering TPDs' needs; otherwise, TPDs might be confused and have difficulties adapting. TPDs discussed various issues when the API version changes from v3 to Kraken (i.e., v5) to Helix. 

First, the previously worked methods might not be supported anymore, and TPDs spend extra effort adapting to the new version. For instance, TPDs can not access Twitch APIs with methods other than the OAuth flow, the authentication flow designated. Furthermore, many features initially supported might still be under Twitch's development roadmap, so TPDs need to figure out approaches to bridge this gap.

Even if TPDs migrated their extensions to the new version, they might feel nervous as they were unsure whether the migration would work. As a TPD shared, \textit{\say{pretty nervous, just pushed out the update that migrates my browser extension from using v3 to using IDs and helix and some v5 fallbacks.}} With the extra workload, TPDs think their innovative momentum has been stopped, especially when TPDs were just introduced to new Twitch tools to help them develop new extensions.
\begin{quote}
    \say{when there have been so many amazing work and progress recently to help developers like me become familiar with developing Extensions and building out better tools like the new Dev Rig for us to just start being able to iterate so much more quickly and efficiently on them and brainstorm greater and more ambitious interactions... And not only that, all the time we've invested into learning how to work with Extensions just so we can start really pushing out code now has to be shifted to fixing up old code. That's why this sucks.}
\end{quote}

Second, TPDs are confused about Twitch API usage and wonder how to use it in what circumstances. Otherwise, the extensions might be banned. For example, a TPD was confused about why they received a warning email from Twitch to stop using v3 (an old API version) because it did not allow TPDs to use v3 APIs anymore. Yet, the TPD thought they had already updated their extensions. Another issue TPDs mentioned is that Twitch did not give them enough time to adapt to the new version. As a TPD complained, \textit{\say{I have the same problem on my public extension. Just not had time since we got told about it two/three months ago to fix it.}}

Without coordination, TPDs used their ways to adapt to the technical updates and keep extensions working. Some of them might fake the new process to skirt the update; as a TPD shared, \textit{\say{the biggest pain point for me moving my bot away from IRC chat commands to Helix API calls was faking the oauth flow once to get a token for it.}}
Some TPDs try to predict Twitch's future updates with some clues they observed and adapt to the prediction in advance. For example, they try not to build extensions with those APIs they guess would not be supported by Twitch anymore, even if there are merits in using these APIs. Some removed the features that they knew would no longer be supported.
While some TPDs took precautions against new updates, some believed the old APIs would last until Twitch created new APIs that could fulfill the exact needs.

TPDs have different opinions about Twitch's technical updates.
Some TPDs think Twitch must improve its infrastructure for future developments, even though that means there would be some costs TPDs need to take.
Yet, some TPDs think the technical updates worsened the Twitch platform and damaged the TPD community.
For example, some TPDs complained that the new changes worsened the experiences of the platform users, such as streamers, especially when the change is drastic.

Moreover, the updates might deteriorate the extension ecosystem by discouraging users from installing extensions. For example, without clicking on the extension, users can only preview an extension via a small icon; a TPD said that \textit{\say{it makes it so much tougher for any of our extensions to get used}}.
Similarly, TPDs complained that the new extension framework made developers further compete with each other over users' usage as Twitch reduced the number of extensions users could use at the same time. As a TPD complained \textit{\say{all the exciting and innovative interactions that Extensions were promised to be are being forced to compete for the single Overlay slot and that's really unfortunate.}}
Some TPDs thought that their issues were not fixed because of the unimportant and little business value of the TPD market on its platform and guessed the reason for the delay and the lack of feedback.

\subsection{Policy and Regulations at Different Levels}

\subsubsection{Platform Policy Update}

TPDs are confused about the platform policy implementation and use Discord as a channel to ask for information about how the policy is implemented or ask for resources about the location of the most updated and recent policy so that they can adjust their expectations about what they should do. Some TPDs are confused about their bot ban because they designed very simple bots following some simple commands with the Twitch policy, but they still got banned. Sometimes, TPDs might try to bypass the limitation of the policy and apply alternatives or some other available features to meet their needs. 
\begin{quote}
    \say{TPD A: Also there are anti-spam measures in place (the specifics of which aren't disclosed). So if you send the same message frequently, or send unsolicited whispers, you may get caught by the anti-spam protections and blocked from whispers. 
    
    TPD B: I mean is it possible, for example sending one thousand whispers maybe in 1000 separate days?}
\end{quote}
In this case, sending whisper messages with a bot might trigger the anti-spam measures, so the TPD considered another way to avoid the measure (e.g., in 1000 separate days). Some TPDs feel the policy enforcement is unfair and would be against the rules, though they already know the violation: \textit{\say{I definitely do something that goes against their rules (logging chat messages from channels). But at the same time so does streamcharts, and they've been around for years.}}

TPDs complained about the Twitch bot review process, which lacked transparency and stability, so no one knew how the mechanism worked. A TPD explained that even though the bot might be verified successfully, it could still be blocked if it triggered Twitch's anti-spam measure and highlighted the power Twitch had to make the final decisions: \textit{\say{You may get verified, you may not and there's no telling how long it may take for a decision. But don't put your hopes on it.}}
The uncertain review process led to folk theories and rumors about how the review process worked or the platform’s intentions regarding TPDs. As a TPD said, \textit{\say{guess it's like two people doing those reviews... Guess extensions don't make so much money compared to the rest.}} The frustration might make TPDs think Twitch does not want to have TPDs develop extensions, as a feeling a TPD shared, \textit{\say{why else would they give us such a hard time.}} Some TPDs considered the potential difficulties of the Twitch side and thought it was unstable due to many factors. As a TPD said, \textit{\say{usually around 2 working days. Then it will depend of your extension and its complexity, how the review team can test it, how many extension have to be reviewed, etc.}}

TPDs also discussed the boundary of Twitch policy regulation and can’t control external extensions from other platforms like Chrome extension. Since different platforms had different policies, some platforms might approve the design, but Twitch would finally reject it. A TPD said, \textit{\say{IGDB will allow images that \_don't quite\_ meet the requirements, but the Twitch sync will reject images that don't \_quite\_ meet the requirements.}} The need to use different platforms for bot development and the inconsistency of policy enforcement cause disruption and extra labor for TPDs.

TPDs asked about copyright issues and the protection of intelligent property. The property protection issue is from both the TPDs' and platform sides. On one side, TPDs wonder if they have the power to control the ownership of their extensions and design ideas and have concerns that Twitch might take their bots with relevant benefits.  For example, a TPD asked, \textit{\say{okay @name and... do I have full control on my extension? like it's my right? like twitch not going to steal it or take credit for it, and twitch in away would show that I made it and it is my and even to a certain extent provide legal protection.}}  Someone is even afraid and joked that they should have a lawyer for a personal project because of the confusing policy regarding using content with explicit permission from the content creator. Conversely, some TPDs also questioned whether they could use Twitch’s property for their bot development, such as using Twitch emotes for their extension and `Twitch' in their bot names. A TPD specifically mentioned that they developed an app for WatchOS and asked whether they call it \textit{\say{Twitch Chat for WatchOS}} and then realized that it is not allowed. Otherwise, they might face lawsuits.

\subsubsection{Data Privacy and Policy}

TPDs need to access data for their apps to function properly while being careful about Twitch's data privacy and policy in terms of services (TOS) and copyright law (the Digital Millennium Copyright Act, DMCA). If bots store unpermitted content (e.g., chat history without permission), Twitch can ban them.

Third-party apps need permission from streamers or viewers to access data on various occasions. For example, when they want to store data for their viewers, they need to identify viewers at first. ID share is useful for an extension to identify and store user data and support the work of an extension. Similarly, if TPDs want to get data from a channel, such as subscription data, they need to make efforts to contact the broadcaster for their permission. However, TPDs might think asking permission for data access is difficult for extension development. Some useful functions might require personal information that users are unwilling to share. For example, TPDs think viewers might not want to share email addresses even if it is the only reliable way for third-party apps to communicate with viewers. 

TPDs might breach data policy due to various reasons. First, TPDs might be confused about what scenarios about storing user-generated content violate Twitch’s TOS. Sometimes Twitch provides data access APIs without preventing TPDs from using them accidentally as a TPD warned another TPD, \textit{\say{Just because the API may technically provide a means to certain data, does NOT mean you can use that data however you choose}}. Providing data access APIs does not mean that data access is legal, yet some TPDs might not be aware of it.  Sometimes, TPDs think they own the data, such as chat logs in their own channels, but they are unsure whether it violates data policy. Second, TPDs might think Twitch would not punish them because Twitch would not find their policy violation, they are non-profit, or they have only a small user base.
 
However, other TPDs might behave more cautiously. A TPD suggested their peers take Twitch policy seriously, as there are several reasons why Twitch has not taken action against the apps that have broken the data policy.
\begin{quote}
    \say{Just keep in mind, the legal process is slow so just because you haven't gotten a Cease and Desist yet doesn't mean they aren't tracking activity and logging things for potential legal action at some point down the line. Secondly, even if you're not making money or you have a small userbase, Twitch can and have sent C\&D letters and shut down small sites in the past. Finally, you have no clue what legal agreements others may have with Twitch. Some sites do enter into agreements with Twitch to use the API in certain ways. So it's pointless to say ``well they do it!'' when you have no clue if they have permissions from Twitch or not.}
\end{quote}
In this case, the TPD reasoned why they suggested not to break Twitch policies. It is because it is possible that Twitch acts slowly or different apps have different legal arrangements with Twitch.

Even if Twitch and the copyright law protect data, some TPDs can still access data without permission from Twitch users.
First, TPDs might consider getting the user’s permission from another third-party app. It is unclear if users are aware of the share of permission. A TPD suggested their peers, \textit{\say{... interact with another 3rd party application that geolocates users when you have permission to do so.}}
Second, data leaks can result from the trust between stakeholders, such as streamers, moderators, and TPDs. TPDs might be delegated by a moderator without the permission of the streamer: \textit{\say{Also a streamer may trust their mods, but mods can delegate ban functionality to 3rd party apps unbeknownst to the streamer.}}
Third, TPDs might omit Twitch’s data regulation and hide their misconduct. Neither Twitch nor Twitch users could avoid data breaches. Someone said, \textit{\say{The bots that are worse such as those that log chat messages don't even need to be logged in to chat, so you'll never see them in chat and have no way to prevent it.}}

After TPDs get permission to access and store user data, TPDs could use Twitch services to validate their users before performing data access from databases. Yet, TPDs might leak data if they do not know how to keep the key to data access safe, as the process of validating users is complicated. For example, if TPDs do not know that they need to keep the authorization token and the refresh token safe, they might unknowingly leak data.

\subsubsection{Discord Community Policy}

TPDs think this Discord is supposed to be an inclusive space for various kinds of people, including TPDs with different levels of knowledge, streamers, and even viewers who want to discuss Twitch-related tool development.
\begin{quote}
    \say{This is a community for everyone on Twitch. TwitchDev is the main focus but its more about creating and building tools for everyone. So broadcasters and viewers are definitely welcome to discuss the future of the platform in terms of tools.}
\end{quote}

A few expectations and rules make the environment comfortable for everyone. 
TPDs expect people to have explored the questions before asking for help in Discord, as a TPD said while answering a question, \textit{\say{I'd assumed you'd already looked in the docs before coming here}}. 
Besides expectations, this Discord server also sets its own rules. 
First, people should not directly message someone without permission to prevent people from getting bothered. They especially think it is important not to message Twitch staff without their agreement directly, and disturbing Twitch staff might worsen the messengers' situations. 
Second, people are allowed to post recruitment messages. Yet, they are not allowed to share their CVs or seek developer jobs or other roles, such as moderators, because it would disrupt the interaction in the community. 
Third, due to the slow reply speed of Twitch support and the lack of legal support from Twitch, Some TPDs resort to this Discord server for Twitch support or legal advice because some Twitch staff are here. Yet, the community prefers not to provide advice as they are not experts; as a TPD said, \textit{\say{Well, some staff here. But this is not a Twitch support server nor a space to get legal advice from.}} 
Fourth, legal questions are commonly asked, but TPDs think it is inappropriate to provide advice for legal questions because they are not lawyers. Different people may interpret the law differently. Some TPDs would share their experiences addressing legal issues; when they do so, they disclaim that their sharing is not legal advice. 

\subsubsection{Bot/Extension Design Rules}

To design Twitch extensions and bots, TPDs must be familiar with complicated Twitch services and understand how to interact with them. An easy way for TPDs to interact with Twitch services is by following Twitch’s design patterns. Yet, TPDs are confused about a specific function's affordances, attributes, or constraints. Confusion about Twitch-provided functions can give TPDs different ideas about how to use Twitch resources to design a bot. Take the Whisper function as an example; one party suggested their peers not to use the Whisper function because it was not reliable, according to their experiences. Yet, the other party said the Whisper function worked fine in their bot systems. The suggestion differences might result from understanding the prerequisites for using the Whisper function, such as passing Twitch's spam filter.

TPDs also questioned the extension development involving compatibility, dependency, and adaption on other platforms, such as extensions for the various browsers and mobile systems. When they develop Twitch extensions on other platforms, they also need to comply with the rules of other platforms. For instance, a TPD said they need Apple Developer Program members to develop extensions on iOS Twitch apps. The compatibility of developing tools was also a concern, as it can be the source of errors. To develop a Twitch extension, TPDs had to use multiple tools; as a TPD shared, \textit{\say{I tried to use the websocket addon to control it, as we use BlackMagic kit for production, along with CasparCG and VMix, so being able to automate certain things would be huge.}} TPDs found it difficult to manage the tools and resources. As a TPD shared, \textit{\say{Yeah I've been working with React on the Extension I'm doing and yeah creating the files was easy. It was everything after that and having to deal with 3 tools being thrown at me along with it.}}

TPDs discussed various kinds of risks and costs while developing Twitch extensions. For example, as an ecosystem, TPDs-made tools aim to support development. Yet, TPDs should evaluate if using non-Twitch-made tools would cost extra effort or contain hidden risks. Furthermore, TPDs discussed risks associated with security, such as the loss of secrets (i.e., keys to data access generated by Twitch) and data leaks to non-Twitch organizations. TPDs also consider operation and maintenance costs. Some developers reduced the costs by choosing cheaper but compromised solutions or closing their services when paying more for server operations was no longer affordable. TPDs discussed various options to transfer the costs to the users or other TPDs, such as allowing other TPDs to use it for free at the beginning and charging a fee after they are used to the tools, and making extension open source so that streamers can use to create their version for their channels. 

Bot design should also consider users' perceptions and experiences. For example, TPDs suggest peers not use the Whisper function because users do not have a good experience interacting with it since spammers abused the function. Broadcasters might hesitate to allow an extension to send chat messages on their behalf. There is a need to balance viewers and broadcasters' agencies regarding using an extension. TPDs need to consider efficiency and performance when designing an extension. They need to consider the worst-case scenarios and think if they can and how to handle them. 

\subsubsection{Complain and Initiate Policy and Platform Changes}

TPDs discussed actions they can take to support better extension development.
On the one hand, TPDs can voluntarily have a consistent and coherent pattern, such as using a unified extension repository to support better extension searchability.
\begin{quote}
    \say{...the ``best'' way to simplify plugin installation is for us to have a remote plugin repository that users can download right in the program unfortunately. I've discussed this in detail with my other guys and that's really the sort of route we'd end up having to go for it to be friendly towards users and satisfy the ``plugin discovery'' needs of plugin creators}
\end{quote}
They also advocate for their peers to make extensions more friendly and accessible. For example, some TPDs call for awareness for streamers with disabilities (e.g., streamers with fibromyalgia).
\begin{quote}
    \say{It's a great topic to bring up though, a couple of my good friends who stream have fibro, one of which was even on a Twitchcon panel on streamers with disabilities and they are helping try to create discussion around this themselves. I'm not sure where the best place to have such a discussion would be, but it's certainly something that needs to happen, and continue to happen to raise awareness.}
\end{quote}
On the other hand, TPDs could use UserVoice (a Twitch platform that allows people to provide design ideas to Twitch) to suggest Twitch make their UI more friendly for extension use. TPDs use Discord to test water if other TPDs would support their suggestions. They also promote suggestions and ask people to upvote: \textit{\say{Maybe, it is better to start small. I'm planning on creating a uservoice to introduce public reviews for extensions from streamers who have installed the extension? Would you guys support something like that?}}
In particular, they mentioned that they think Twitch could consider changing the petition if a significant number of developers are advocating it. They also consider asking streamers to vote for their petitions, especially those they work for.

\subsection{Seeking and Showing Support}
\subsubsection{Information and Instrumental Support}

TPDs seek various information support in Discord because TPDs might not find good tutorials, documents, or official tech support when they want to develop Twitch third-party extensions. 
For example, new TPDs seek advice for learning paths, tips, examples, and experiences for Twitch extension development, such as the languages to use, what kinds of projects are good to start with, and documentation to start quickly.

Some of the new extension developers are not new to programming but new to Twitch extension development. They seek templates or packages to guide them to pick up the developing patterns quickly. For example, a bot developer looked for methods to make their bot programs work as extensions. TPDs also seek practical development advice, such as analysis about choice for developing tools, efficient algorithms, feedback on current development plans, and suggestions for debugging. Usually, they provide detailed descriptions for the peer TPDs to understand the problems they encounter. Besides programming suggestions, TPDs expressed the need to have a proper tool kit to actualize certain design ideas:
\textit{\say{With a proper SDK I can try some of the ideas I had a while back for working with Blackmagic switchers/ATEM, in conjunction with our VT setup (either VMix or CasparCG depending on the situation).}}

There are a few roles TPDs want to seek support from. Usually, people ask the whole Discord server questions and appreciate any help. Besides asking questions in the Discord server, people could also look up official posts or consult commercial third-party agencies specializing in developing Twitch-related programs. However, they sometimes need help from specific roles, such as Twitch engineers, Twitch extension review staff, and people with legal expertise. 

In general, TPDs are willing to help their peers. TPDs respond to support requests in various ways. For example, when there are already well-written or well-organized external resources suitable for answering the questions, TPDs redirect people to these, such as websites, blogs, or commercial third-party supportive agencies. TPDs also pointed out the malpractices or shared failure experiences to prevent peers from being trapped. Some TPDs provided instrumental support to benefit the TPD community. For instance, a TPD mentioned that they tried to build foundational programs that people could learn from and build upon. On the other hand, the Twitch DX team also actively looks for feedback from TPDs regarding Twitch UI, documentation, and products. However, there are some problems this Discord server could not help with, such as helping TPDs get a quicker response from Twitch staff.

\subsubsection{Socializing with other TPDs}
TPDs are glad that there is a Discord server for TPDs. Compared with the official Twitch developer forum, this Discord server serves as an interactive panel small enough that TPDs could feel included and talk more. TPDs here ask for or provide help and encourage their peers, share an appreciation for help, chat about their lives and leisure, update their developing status, and share personal thoughts. Furthermore, they share their experiences in Twitch meet-ups or Twitch-related events and discuss attending events, such as Twitch AWS Extension Challenge, Twitchcon, and extension workshops.

\section{Discussion}
In this study, we redirected our attention from the tool user's perspective and the scholarly focus on tools' social and functional roles to TPDs engaged in developing tools for community management. We highlighted the importance of TPDs as critical stakeholders and a distinct user group in community management. These findings can potentially guide future work related to novel platforms with a high demand for TPDs to supplement their platform ecosystems, such as TPDS for VR  and generative AI platforms, as these platforms quickly evolve with advanced technological infrastructures. 
We confirmed that TPDs for tool development align with independent developers in other domains, such as gaming \cite{freeman_understanding_2023}. This consistency underscores how the emerging open and participatory production model is instrumental in forming new workforces and fostering innovative technological practices within the worldwide technology industry \cite{flecker2010organisational}. While the ecosystems with open-source development models foster innovation and creativity, the tensions and challenges regarding policy and power with platforms and different stakeholders restrict TPDs' agency to actualize innovation and creation.  

\subsection{Co-Creation for Idea Innovation with Tool Users}
Prior work suggests that regular meetings with end-users helped to keep the software developer aware of the usability without having to devote extensive periods to it \cite{prior_use_2013}. Engaging with streamers and viewers allows TPDs to gain insights into user needs, pain points, and preferences, resulting in tools that align more closely with user expectations. This collaborative approach can lead to innovative solutions that might not have been possible through TPDs' efforts alone. The Discord community mediates co-creation, idea innovation, and support with tool users, motivating TPDs' autonomy to develop practical tools to benefit the ecosystem.   

The Discord community is a space that Twitch staff, Twitch users, and TPDs are all familiar with, as it inherits the community of Twitch streamers and their fans, and some TPDs are also streamers. It serves as a third space for stakeholder gathering.
This space allows for exchanging ideas, feedback, and innovation, leading to improved tool development and enhanced engagement with the user community, as shown in prior work \cite{kiene_technological_2019}. Discord's platform facilitates direct and real-time communication (unlike thread-like comments on Twitter and Reddit, more like chronological chat with optional voice communication \cite{jiang_moderation_2019}), enabling TPDs to stay connected with users and receive continuous input. In this sense, Discord, as a space, lowers the barrier for diverse stakeholders, facilitates co-creation, and results in a win-win situation for multiple stakeholders. 
This indicates the importance of considering not just the front-end users and the core platform in understanding the ecosystem's landscape but also those supporting the platform and other spaces facilitating interactions among stakeholders behind the scenes.

The legal and privacy concerns impose restrictions on obtaining tool users' data,  limiting their ability to understand users' needs and impeding the actualization of these designs. Prior work shows that developers might request multiple permissions because of the confusion of permission scope and policy requirement {\cite{tahaei_stuck_2023}}. The data permission and legal regulation intertwined with Twitch platform policy confuse TPDs and sometimes even cause violations and ethical issues of tool development.
How to balance offering user data permissions to stimulate design innovation and protecting users' privacy requires further investigation.

\subsection{Twitch Infrastructure Update and Policy Dynamics: Conflict of Control}
The relationship between TPDs and Twitch itself, particularly regarding technical updates, policy changes, and design rules, can create complexities in TPDs' efforts. Like independent developers who often lack the financial, social, and technical support to sustain their ongoing labor and production \cite{freeman_understanding_2023}, TPDs lack socio-technical and political support from the platform, restricting their autonomy. These updates can directly impact their tool performance and their psychological well-being.

Platform's function and interface update can significantly broaden the development knowledge gap between the platform and third parties and increase TPDs' barriers to understanding and aligning their development and innovation with the platform's vision \cite{foerderer_knowledge_2019}. 
TPDs must quickly adapt their tools to frequent technical updates and evolving policies on  Twitch. Evolving policies can impose a significant administrative burden, necessitating additional labor to stay informed and prevent operational disruptions. TPDs must continually monitor and adjust to these policy changes to maintain compliance. The lack of clarity in these policies leads to rumors and confusion among TPDs, impacting their decision-making and, potentially, their financial stability.

As a leading live streaming platform in this industry, Twitch can do whatever it wants, such as keep updating policies and infrastructure, causing much labor and confusion to the TPDs, who are valuable stakeholders of the platform ecosystem. Dangerously, many TPDs might act against the policy by circumventing the policy regulation when they feel it is unfair and might harm their communities. This could potentially deteriorate the ecosystem and drive TPDs away if they can not get their bot approved in time (e.g., they had tried many times to keep updating their bots),  which takes more physical and physiological labor. Furthermore, clarifying these policies and disentangling them to simplify policy presentation while ensuring they are up-to-date and transparent remains challenging. This is especially true given the rapid pace of development and technological advancements in the industry, warranting further investigation.

Prior work has shown the platform crafted the boundary resources (the tools and regulations used to maintain the relationship between the platform and TPDs) for cultivating the platform ecosystem \cite{ghazawneh_balancing_2013}. While Twitch provides the technical resources to empower TPDs' agency, it also provides multi-layered policies to regulate TPDs' autonomy. In our case, policies might have a more substantial impact than the technical resources, reflecting the platform's conflict of control. On the one hand, it hopes to embrace and help the platform's ecosystem flourish with open API and tool kits; on the other hand, it reinforces multi-layered policies at different development stages, from developing, reviewing, and deploying, to updating their tools. While TPDs use their boundary resources to build complementary applications, the development is a knowledge-intensive task that crosses organizational boundaries \cite{kauschinger_detecting_2023}.

Building on previous research in HCI focused on technological design to enhance developers' skills and productivity (e.g., \cite{guzzi_supporting_2015,mellis_collaboration_2012,li_exploring_2023, li_channeling_2022}), we propose that future studies should incorporate an understanding of policy dynamics that influence TPDs' development capabilities within the socio-technical system design. This approach would integrate policy into the technological framework, providing a more holistic view of the factors that impact TPDs' effectiveness and efficiency. 

We clarify that we don't want to criticize the platform, as an evolving and emerging technology \cite{hamilton_streaming_2014}, it shows its effort to improve its systems for its stakeholders, such as developing new tools, enriching existing tool features, and working on transparency report \cite{nightingale_twitch_2022}. Updating and changing are good ways to improve the platform's user experience. However, it might cause harm to the ecosystem (e.g., TPDs felt they were not valued with all the barriers they encountered). Future work should explore how to improve the ecosystem's health with effective governance mechanisms \cite{fontao_supporting_2018} such as mitigating potential policy violations and balancing the empowerment and control, and ensuring that all stakeholders are synced and avoiding the extra labor in communication, coordination, and development.

\subsection{TPDs and Community Governance}

Different from independent game developers who often work as a small team \cite{freeman_understanding_2023} and can publish products in alternative ways such as self-funding and small studios \cite{freeman_mitigating_2020}, TPDs primarily work as individuals for tool development and have to rely on the platforms' governance politically and technically.  TPDs take the semi-private space to form a community, exchange and innovate design ideas,  socialize with peers, and show information and instrumental support regarding the platform's policy, technical updates, and potential ways to circumvent the policies and their lessons. As a reflection of constructive suggestions, these supports maintain their peers in the community \cite{huang_effectiveness_2016}.

TPDs also build community guidelines to facilitate constructive interactions.  However, it might also cause issues regarding interactions between different stakeholders.  We explicitly see evidence that Twitch developers welcome interaction, whereas the Discord community rules explicitly mention that they should not be bothered without permission unless there is an emergency. Defining a \textit{\say{emergency}} issue in the community is unknown. The community guidelines seem to deter TPDs and demotivate the Twitch developers to engage.

Twitch's vague and inconsistent policies and regulations also motivate TPDs to initialize changes regarding tool development, with the collaboration with other stakeholders (e.g., streamers) to improve their voice and power to negotiate with Twitch, a good indicator of external developers to unionize and restructure power dynamics \cite{ruffino_independent_2020, freeman_mitigating_2020}.   That is a good start, but they also faced many challenges regarding the coordination of the community and the structure and governance of their communities. It seems like there is a clear power structure in this community (admins, Twitch developers, TPDs, and tool users). A core concern is how to effectively govern online communities while not demotivating multiple stakeholders.  How to maintain the community can also be complex. In this study, we only know they started the idea to work with a formal community to unify the bot development for their community. We know little about the work on this process, which needs further inquiry to support their community governance, with questions like what structure helps them thrive.

\subsection{Design Implications}

While TPDs shared positive experiences regarding idea innovation and support from the community, they also faced challenges from the Discord community governance and Twitch's technical and policy updates. We advocate for preserving the positive aspects of user experience while mitigating the negatives. Therefore, we propose designs that sustain innovation on Discord and enhance idea actualization on Twitch.

\subsubsection{Support the Ecosystem by Facilitating Collaboration and Maintaining Idea Innovation on Discord}

As the Discord community is the hub for communication and collaboration among stakeholders, we propose design implications to maintain idea innovation and facilitate stakeholder collaboration.

\paragraph{Maintain the co-creation and innovation among TPD and tool users}
Though we found that Discord, as a third space, provides opportunities for stakeholders to exchange and innovate design ideas and that TPDs and peer learning thought conflict are minimal, and information and instrument support are abundant, we also identified some issues that would improve the community innovation if these are fixed, e.g., to integrate the Twitch forum and r/reddit into the communities. Prior work shows that some TPDs actively interact with streamers in the Twitch subreddit (r/Twitch) to understand their needs and challenges of their community management and moderation and develop a bot for them \cite{cai_hate_2023,han_hate_2023}. Prior work shows developers leverage API to develop personalized bots for their core platform for collaborative purposes {\cite{lin_why_2016}}. Similarly, TPDs and Discord designers might consider utilizing the Discord API to create bots that facilitate community maintenance, addressing specific needs identified by TPDs, such as identifying and summarizing the design ideas from various discussion {\cite{zhang_making_2018}} of tool users (e.g., streamers, moderators, viewers). Such an initiative would allow TPDs to synchronize all relevant information within the Twitch ecosystem, potentially leading to more effective co-creation for community management.

\paragraph{Facilitate TPDs and Twitch developer collaboration} 
Twitch developers might work as an oversight board but don't engage much about Twitch's most recent updates. The ineffective communication provides little support to TPDs. Prior work suggests that professional developers should maintain control and flexibility when interacting with open-source project developers \cite{wissel_how_2019}. We don't know whether Twitch developers face similar challenges when interacting with TPDs because of the platform policy and terms of service.  However, we can still argue that the Twitch developers in the community should be more flexible and engaging under the Twitch terms of service. Improving communication and collaboration channels between TPDs and Twitch developers is crucial to address these challenges. This could involve creating more direct lines of communication and providing more explicit guidance on policy and technical changes. For example,  the Discord community can leverage the social roles of bots to integrate a voting system with a notification mechanism to remind a certain role in the community (e.g., Twitch staff) to intervene; such a way can also avoid the disturbing effect of frequent messaging from the community. When an idea is voted, it will stick to the top of the channel; when the vote achieves a certain number (e.g., 20 or 40), the system will send a reminder to all Twitch developers. Then, someone can intervene with a message or voice chat to clarify everything. This is a more socialized design with multi-modal interaction and real-time feedback within the TPDs' community, different from the Twitch UserVoice voting system, which is more centralized and static.

\subsubsection{Unify the Development Tools and Policies to Clarify the Development Workflow on Twitch}

According to the boundary resource model {\cite{ghazawneh_balancing_2013}}, Twitch might overly control the development process with technical and policy updates, restricting TPDs' autonomy; thus, the platform should provide more resources to empower TPD's agency and facilitate the idea actualization.  Establishing mechanisms for TPDs to provide feedback on Twitch updates and policies can help bridge the gap between the platform and TPDs.
 
\paragraph{Unify the toolkit for development}

Though Twitch has a tool kit for TPDs, many TPDs share their development practices with different programming tools. They faced debugging challenges, meaning that TPDs might need an enriched and unified tool kit to facilitate their development. Therefore, we suggest that the platform should expand and standardize the existing toolkits, incorporating feedback from the TPDs, as supported by Terry et al.'s research on the importance of collective input \cite{terry_perceptions_2010}.
One possible approach is for the platform to create a matrix categorizing tool sections guided by bot/extension design principles and user needs identified in our research. This matrix would allow TPDs to contribute tools they have found helpful in development. Other TPDs could then vote on or add to these suggestions, ultimately leading to a consensus on a set of tools for each category. The finalized toolkit, enriched by this collaborative process, could then be integrated into the platform ecosystem, complete with relevant resources to support TPDs more effectively. However, the scale of the tool sections and the broad policy alignment among these tools might cause extra challenges for TPDs, warranting further exploration. 

\paragraph{Clarify the policy kit from development}
The inconsistent and opaque platform policy compounding legal issues limited TPDs' development autonomy. Though Twitch provides more than technical documentation such as blog posts and online communities (e.g., Twitch UserVoice), these are designated as self-service and have the potential to scale up like helpdesks \cite{foerderer_knowledge_2019, kauschinger_knowledge_2021}. 
Prior work has shown that policy development can be categorized into a hierarchy to govern community \cite{chandrasekharan_internets_2018}. We propose a policy kit to hierarchically structure the rules to guide novice and experienced TPDs. The policy tool kit provides professional developers and TPDs access to update relevant parts with one or two administrators to approve or decline the edit, similar to the crowd documentation \cite{parnin_measuring_2011}, but in a structured format about policy and other necessary documentation, with the involvement of the direct stakeholders to articulate the procedure of governance \cite{zhang_policykit_2020}. 

\subsection{Limitations and Future Work}
There are several limitations. First, this is a preliminary study of TPDs' tool development. However, we only looked at the data in the main channel (``Lobby''). Other channels might have a more in-depth discussion of a specific category. Future work should extend this work and answer questions like how the TPDs build a network offline and how offline interaction affects their collaboration and innovation in the online space. Second, the qualitative study reveals the categories, yet we don't know how the events affect the interaction dynamics in this community. Future work can extend this study by running quantitative analysis to understand the interaction dynamics with community growth. Third, the findings mainly showed the policy and stakeholders' impact on tool development. We only have limited evidence about the tool development process. Future work can run qualitative studies, such as interviewing and observing TPDs in this community, to understand their design in-depth, such as the mental model of tool development. Fourth, our study is limited to Twitch's TPDs; however, TPDs across various live streaming platforms may operate within distinct hierarchies and political cultures, such as those in China. Future research should broaden its scope to include live streaming platforms across diverse contexts, aiming for a more comprehensive understanding of TPDs' practices. Lastly, the topic modeling approach in our study, focusing on two selected topics, may overlook nuanced details of other aspects. Future studies should employ more advanced techniques for topic filtration to fully capture TPDs' practices.

\section{Conclusion}

In this study, we delve into the practices and challenges faced by third-party developers (TPDs) within the Twitch live streaming platform, who play a vital role in mediating community management with tool development. Analyzing discussions in a semi-private TPD community on Discord, the study reveals that TPDs contend with intricate policy regulations that influence their tool development. Additionally, they serve as intermediaries between diverse stakeholders, including streamers, viewers, the platform, and professional developers. Their ability to manage these complex relationships and policies is crucial to strengthening their autonomy and innovation capabilities and actualizing tool development. 
This study contributes to the existing body of research on online community management by offering a unique perspective from the TPD's point of view. We underscore the importance of supporting their autonomy and agency in the development process. By recognizing the challenges and complexities they face, we can work towards creating a more effective and user-friendly ecosystem for online community management.
\balance

\bibliographystyle{ACM-Reference-Format}
\bibliography{TPD,sample-base}

\appendix
\section{Thematic Analysis with Codebook Definition}
\label{apx:codebook}

\begin{table*}[ht]
\resizebox{\textwidth}{!}{%
\begin{tabular}{lllll}
\hline
Theme &
  Sub-themes &
   &
  Code &
  Definition \\ \hline
\multirow{8}{*}{\begin{tabular}[c]{@{}l@{}}Design ideas exchange \\ and innovation\end{tabular}} &
  \multirow{4}{*}{\begin{tabular}[c]{@{}l@{}}Streamer's needs\\ and design ideas\end{tabular}} &
  22 &
  streamer can't work for niche features &
  \begin{tabular}[c]{@{}l@{}}Streamers do not know how to implement personalized features for their channels, \\ and they require TPDs' help.\end{tabular} \\
 &
   &
  24 &
  streamer's idea about designing fun VIP status &
  Streamers need features that can enhance viewers' engagement and experiences. \\
 &
   &
  35 &
  streamer: whisper or email as messaging tool &
  \begin{tabular}[c]{@{}l@{}}Streamers are confused about tool development related to communicating with their \\ viewers.\end{tabular} \\
 &
   &
  41 &
  streamer's idea about the audible indicator feature &
  \begin{tabular}[c]{@{}l@{}}Streamers' needs related to streaming notification control, such as the ways to stop \\ notifying them whenever a viewer leaves.\end{tabular} \\ \cline{2-5} 
 &
  \multirow{4}{*}{\begin{tabular}[c]{@{}l@{}}Exchange design\\  ideas\end{tabular}} &
  6 &
  first extension showcase &
  \begin{tabular}[c]{@{}l@{}}TPDs showcase their first Twitch extension, including descriptions and testing \\ or demo links.\end{tabular} \\
 &
   &
  14 &
  first extension design idea &
  First tool development ideas TPDs share. \\
 &
   &
  33 &
  first extension practice &
  TPDs share the practices about how they worked on their first extension. \\
 &
   &
  49 &
  general design extension idea &
  Tool, bot, or extension design ideas TPDs share. \\ \hline
\multirow{11}{*}{\begin{tabular}[c]{@{}l@{}}Impact of Twitch \\ technical update\end{tabular}} &
  \multirow{2}{*}{Review extension} &
  25 &
  time of approval of extension &
  TPDs want to know when their extensions will be approved. \\
 &
   &
  55 &
  time of review of extension &
  How long does the Twitch staff take to review their extensions? \\ \cline{2-5} 
 &
  \multirow{9}{*}{Technical updates} &
  2 &
  refresh before the expiration of the subbed topic &
  There is confusion about when Twitch decides to expire a subbed topic. \\
 &
   &
  8 &
  refresh token &
  Confusion about when to refresh tokens for apps' permission. \\
 &
   &
  9 &
  change of version &
  TPDs' discussion related to Twich's change of system versions, such as the API version. \\
 &
   &
  13 &
  Twitch boilerplate with React is straightforward &
  Discussions related to Twitch's supportive tools for TPDs, like boilerplate \\
 &
   &
  23 &
  OBS plugin move source &
  Discussions related to plugins from platforms like OBS \\
 &
   &
  28 &
  something like API that can be public &
  something like API that can be public \\
 &
   &
  37 &
  twitch extension challenge &
  Challenges related to extension development \\
 &
   &
  50 &
  extensions availability on different platforms &
  \begin{tabular}[c]{@{}l@{}}TPDs discuss that some extensions are available on some platforms, like Chrome, but not \\ on other platforms, like gaming consoles.\end{tabular} \\
 &
   &
  52 &
  multiple tool use is difficult &
  \begin{tabular}[c]{@{}l@{}}TPDs complain that they find it difficult to simultaneously use multiple tools for\\  extension development\end{tabular} \\ \hline
\multirow{10}{*}{\begin{tabular}[c]{@{}l@{}}Policy and regulations\\  at different levels\end{tabular}} &
  \multirow{2}{*}{\begin{tabular}[c]{@{}l@{}}Platform policy \\ update\end{tabular}} &
  32 &
  Twitch evaluates bots &
  Discussions related to how Twitch evaluates TPDs' bots. \\
 &
   &
  34 &
  extension design rule /consideration/explanation &
  \begin{tabular}[c]{@{}l@{}}TPDs discuss considerations and rules when implementing or designing a certain feature \\ of an extension.\end{tabular} \\ \cline{2-5} 
 &
  \begin{tabular}[c]{@{}l@{}}Data privacy \\ and policy\end{tabular} &
  60 &
  \begin{tabular}[c]{@{}l@{}}relevant but not listed: privacy, \\ security, authentication, permission\end{tabular} &
  Discussions or concerns related to Twitch's privacy policy and access control rules \\ \cline{2-5} 
 &
  \multirow{4}{*}{\begin{tabular}[c]{@{}l@{}}Discord community \\ policy\end{tabular}} &
  39 &
  redirect to a specific discord channel &
  Members redirect people who ask questions that are related to a specific channel. \\
 &
   &
  45 &
  TPD dispute &
  Disputes among TPDs \\
 &
   &
  47 &
  No legal inquiry and advice &
  \begin{tabular}[c]{@{}l@{}}The Discord community moderators and members say that it is inappropriate to consult\\  legal related issues in the Discord server.\end{tabular} \\
 &
   &
  57 &
  community guideline &
  Announcement or discussion about discord community guideline \\ \cline{2-5} 
 &
  Bot design rules &
  34 &
  extension design rule /consideration/explanation &
  \begin{tabular}[c]{@{}l@{}}TPDs discuss considerations and rules when implementing or designing a certain feature\\  of an extension.\end{tabular} \\ \cline{2-5} 
 &
  \multirow{2}{*}{\begin{tabular}[c]{@{}l@{}}Complain and initiate\\ unified rules\end{tabular}} &
  60 &
  relevant but not listed: dev group action &
  TPDs advocate other TPDs to have standard ways regarding tool development and design. \\
 &
   &
  60 &
  relevant but not listed: extension use &
  TPDs advocate tool users to use extensions in better ways. \\ \hline
\multirow{12}{*}{\begin{tabular}[c]{@{}l@{}}Seeking and showing\\ support to different \\ stakeholders\end{tabular}} &
  \multirow{3}{*}{\begin{tabular}[c]{@{}l@{}}Socializing with \\ other TPDs\end{tabular}} &
  1 &
  Extension workshop &
  TPDs talk about the participation of workshops related to Twitch extension development. \\
 &
   &
  3 &
  integration with the dev forum &
  \begin{tabular}[c]{@{}l@{}}TPDs discuss the integration among various platforms for TPDs to discuss Twitch-related \\ development, such as the dev. forum and the Discord channel.\end{tabular} \\
 &
   &
  4 &
  happy to have discord &
  TPDs express their positive feelings about having this Discord channel to interact with other TPDs. \\ \cline{2-5} 
 &
  \multirow{9}{*}{\begin{tabular}[c]{@{}l@{}}Information and \\ instrumental\\  support\end{tabular}} &
  38 &
  ping me to help me &
  TPDs ask others to provide support via private messages or notifications. \\
 &
   &
  40 &
  help test extension &
  TPDs want someone to test their extensions for them. \\
 &
   &
  42 &
  Twitch DX: feedback on dev.twitch.tv &
  Twitch Developer Experience staff seeks feedback from TPDs. \\
 &
   &
  43 &
  personalized bot better than other bots. &
  TPDs discuss personalized bots \\
 &
   &
  44 &
  aware of coder error by comparing with others &
  \begin{tabular}[c]{@{}l@{}}TPDs are aware of their errors when they compare the outcomes or processes \\ of other TPDs with theirs.\end{tabular} \\
 &
   &
  48 &
  web app for mods to modify bot &
  TPDs create web interfaces for moderators to conveniently modify bots. \\
 &
   &
  51 &
  TPD facilitates peer learning &
  TPDs act to support the learning of their peers, such as by sharing their coding experiences. \\
 &
   &
  53 &
  extension's request cost &
  TPDs consult peers about issues related to the cost of maintaining the extension service. \\
 &
   &
  60 &
  relevant but not listed: seek for collaborators &
  TPDs or streamers seek collaboration with other TPDs. \\ \hline
\multicolumn{5}{l}{\begin{tabular}[c]{@{}l@{}}The following are general information sharing and support (Q\&A) that are common in all online communities and are not explicit about tool development. \\ We just briefly mentioned the findings and social support section at the beginning.\end{tabular}} \\ \hline
\multirow{4}{*}{} &
  \multirow{4}{*}{\begin{tabular}[c]{@{}l@{}}Twitch ID and \\ data access Q\&A\end{tabular}} &
  10 &
  opaque id as key to store data &
  Discussion regarding methods to store and access data \\
 &
   &
  11 &
  ID share for data access &
  Discussion about what Twitch tool to use to request user's permission for data access \\
 &
   &
  15 &
  chat logs record which ID &
  Discussion about chat logs \\
 &
   &
  46 &
  replacement to get chat history &
  Discussion about alternative ways for accessing chat history \\ \cline{2-5} 
\multirow{9}{*}{} &
  \multirow{9}{*}{\begin{tabular}[c]{@{}l@{}}Testing and \\ debugging Q\&A\end{tabular}} &
  5 &
  Reasons to use rig &
  Reasons to use Twitch-provided testing environment \\
 &
   &
  7 &
  browse through the bot log and browser &
  Discussion about accessing extension status \\
 &
   &
  19 &
  app start errors &
  Discussions about errors when launching an extension or an app \\
 &
   &
  20 &
  distro is not debian like a current kernel &
  Discussion about general environment choice \\
 &
   &
  26 &
  Linux for local Redis testing &
  Discussion about environment setting for extension testing \\
 &
   &
  27 &
  the error of creating a new app &
  Discussions about errors when creating a new app \\
 &
   &
  36 &
  testing backend &
  Discussion about how to test at local backend \\
 &
   &
  56 &
  time to sort out bugs &
  Discussion about the timing to address bugs \\
 &
   &
  59 &
  frontend backend error &
  Discussions about frontend or backend errors \\ \cline{2-5} 
\multirow{7}{*}{} &
  \multirow{7}{*}{\begin{tabular}[c]{@{}l@{}}Visual and audio\\  design issues\end{tabular}} &
  16 &
  transcode options for streaming quality &
  Settings for streaming quality \\
 &
   &
  21 &
  red hat logo &
  Discussions about extension logos \\
 &
   &
  29 &
  icons' resolutions &
  Discussions about icon's resolutions \\
 &
   &
  30 &
  transition/animation &
  Discussions about visual transition and animation \\
 &
   &
  31 &
  generate names of items &
  How to generate names of items \\
 &
   &
  54 &
  tips for image testing &
  tips for image test \\
 &
   &
  58 &
  raw message format change &
  How to do format change for raw messages \\ \cline{2-5} 
\multirow{4}{*}{} &
  \multirow{4}{*}{Other} &
  0 &
  Irrelevant &
  TPD's comments that are irrelevant with extension design and development \\
 &
   &
  12 &
  traffic between viewer and broadcaster &
  Discussion related to how Twitch updates affect connection speed. \\
 &
   &
  17 &
  stability is essential for money for streamers &
  Discussion related to internet stability \\
 &
   &
  18 &
  Internet provider &
  Discussion related to internet providers \\ \hline
\end{tabular}%
}
\caption{Themes, sub-themes, codes with definitions.  The developed codebook contains a total of 61 codes (59 primary codes and two functional codes, where code 0 indicates `irrelevant' and code 60 indicates `relevant but not listed')}
\label{tab:codebook}
\end{table*}

\section{Finding Summarization with Description}
\label{apx:findings}
\begin{table*}[ht]
\resizebox{\textwidth}{!}{%
\begin{tabular}{@{}lll@{}}
\toprule
Section &
  Subsection &
  Description \\ \midrule
\multirow{3}{*}{\begin{tabular}[c]{@{}l@{}}Design Ideas Exchange\\  and Innovation\end{tabular}} &
   &
  \begin{tabular}[c]{@{}l@{}}How Twitch Dev. Discord members innovate, share, and give feedback regarding \\ Twitch extension design.\end{tabular} \\ \cmidrule(l){2-3} 
 &
  \begin{tabular}[c]{@{}l@{}}Streamer's Need and \\ Design Ideas\end{tabular} &
  \begin{tabular}[c]{@{}l@{}}*Streamers seek advice on developing extensions that fulfill various streaming needs.\\ *Streamers want extensions that provide rich interactivity to their audience or\\  specialized extensions.\end{tabular} \\ \cmidrule(l){2-3} 
 &
  \begin{tabular}[c]{@{}l@{}}Exchange Design Ideas\\  among TPDs\end{tabular} &
  \begin{tabular}[c]{@{}l@{}}*Categories of extension design ideas shared by TPDs.\\ *TPDs share sandbox versions of bots and seek feedback for improvement.\\ *Experience sharing and exchange of design ideas.\end{tabular} \\ \midrule
\multirow{3}{*}{\begin{tabular}[c]{@{}l@{}}Impact of Twitch \\ Technical update\end{tabular}} &
   &
  \begin{tabular}[c]{@{}l@{}}The change of technical infrastructures and its impact on TPD’s tool development, \\ such as API version change.\end{tabular} \\ \cmidrule(l){2-3} 
 &
  Schedule Uncertainty &
  \begin{tabular}[c]{@{}l@{}}*TPDs have many questions about uncertain review update time.\\ *Unstable review time.\end{tabular} \\ \cmidrule(l){2-3} 
 &
  \begin{tabular}[c]{@{}l@{}}Extra Labor Caused by \\ Technical Updates\end{tabular} &
  \begin{tabular}[c]{@{}l@{}}*Extra efforts are required to adapt technical update.\\ *Technical update leads to potential extension ban.\\ *TPDs’ own ways to adapt technical updates.\\ *Mixed opinions about Twitch’s technical updates.\end{tabular} \\ \midrule
\multirow{6}{*}{\begin{tabular}[c]{@{}l@{}}Policy and Regulations \\ at Different Levels\end{tabular}} &
   &
  \begin{tabular}[c]{@{}l@{}}The policies and rules that various regulators make and how these policies.\\ influence TPDs, such as guidelines about addressing user data or how Discord \\ members interact with different channels.\end{tabular} \\ \cmidrule(l){2-3} 
 &
  Platform Policy Update &
  \begin{tabular}[c]{@{}l@{}}*The vague or opaque regulations would confuse TPDs, and TPDs would adapt this \\ confusion in various ways.\\ *TPDs’ confusion about the platform policy implementation and consequent reactions.\\ *Opaque and unstable bot review rules and the consequences.\\ *Misaligned policy regulation among Twitch and other platforms.\\ *Confusions about the use and ownership of copyrights and intelligent property.\end{tabular} \\ \cmidrule(l){2-3} 
 &
  Data Privacy and Policy &
  \begin{tabular}[c]{@{}l@{}}*User's permission for data access.\\ *Reasons for data policy breach.\\ *Access data without permission from Twitch users.\end{tabular} \\ \cmidrule(l){2-3} 
 &
  Discord Community Policy &
  Expectations and rules. \\ \cmidrule(l){2-3} 
 &
  Bot/Extension Design Rules &
  \begin{tabular}[c]{@{}l@{}}*TPDs’ confusions about Twitch’s vague design patterns and principles.\\ *Compatibility, dependency, and adaption on other platforms.\\ *Risks and costs.\\ *Users' perceptions and experiences.\end{tabular} \\ \cmidrule(l){2-3} 
 &
  \begin{tabular}[c]{@{}l@{}}Complain and Initiate Policy \\ and Platform Changes\end{tabular} &
  \begin{tabular}[c]{@{}l@{}}*TPDs use Discord for their voluntarily consistent and coherent pattern of extension.\\ *TPDs use Discord to unify their suggestions to Twitch.\end{tabular} \\ \midrule
\multirow{3}{*}{Seeking and Showing Support} &
   &
  \begin{tabular}[c]{@{}l@{}}What kinds of support do people request in Discord, and how people seek\\  and provide support.\end{tabular} \\ \cmidrule(l){2-3} 
 &
  \begin{tabular}[c]{@{}l@{}}Information and Instrumental \\ Support\end{tabular} &
  \begin{tabular}[c]{@{}l@{}}*Seek community support due to a lack of official support.\\ *Support needs from new extension developers.\\ *Roles TPDs seek support from.\\ *Ways TPDs respond to support requests.\end{tabular} \\ \cmidrule(l){2-3} 
 &
  Socializing with other TPDs &
  Discord as a smaller and more inclusive space for TPDs. \\ \bottomrule
\end{tabular}%
}
\caption{Results summary }
\label{tab:findings}
\end{table*}

\end{document}